\documentclass[preprint,pre]{revtex4}
\usepackage[T1]{fontenc}
\usepackage[latin1]{inputenc}
\usepackage[dvips]{graphicx}
\usepackage{float}
\usepackage{amssymb}
\usepackage{amsmath}

\newcommand{\lav}{\left\langle}
\newcommand{\rav}{\right\rangle}

\restylefloat{figure}

\begin{document}
\title{Amplification of Molecular Traffic Control in catalytic grains with novel channel topology design}
\author{Andreas Brzank\textsuperscript{1,2}, Gunter M. Schütz\textsuperscript{1}}
\affiliation{\textsuperscript{1}Institut f\"ur Festk\"orperforschung, Forschungszentrum J\"ulich,
52425 J\"ulich, Germany\\
\textsuperscript{2}Fakultät für Physik und Geowissenschaften, Universität Leipzig, Abteilung Grenzflächenphysik,
Linnestrasse 5, D-04103 Leipzig, Germany }

\email{a.brzank@fz-juelich.de}
\date{\today}

\begin{abstract}
We investigate the conditions for reactivity enhancement of catalytic
processes in porous solids by use of molecular traffic control (MTC). 
With dynamic Monte-Carlo simulations and continuous-time master 
equation theory applied to the high concentration regime
we obtain a quantitative description of the MTC effect for a network of intersecting
single-file channels in a wide range of grain parameters and for optimal external
operating conditions. Implementing the concept of MTC in models with specially designed 
alternating bimodal channels we find the efficiency ratio (compared with a topologically 
and structurally similar reference system without MTC) to be enhanced with increasing 
grain diameter, a property verified for the first time for an MTC system. Even for short 
intersection channels, MTC leads to a reactivity enhancement of up to approximately 
65\%. This suggests that MTC may significantly enhance the efficiency of a catalytic 
process for small as well as large porous particles with a suitably chosen binary channel topology.
\end{abstract}
\maketitle

\section{Introduction}

Zeolites are used for catalytic processes in a variety of applications, e.g.
cracking of large hydrocarbon molecules. In a number of zeolites diffusive
transport occurs along quasi-one-dimensional channels which do not allow guest
molecules to pass each other \cite{Baer01}. Due to mutual blockage of reactants
$A$ and product molecules $B$ under such {\it single-file conditions}
\cite{Karg92} the effective reactivity of a catalytic process $A\to B$ -- determined
by the residence time of molecules in the zeolite -- may be considerably reduced
as compared to the reactivity in the absence of single-file behaviour.
It has been suggested that the single-file effect may be circumvented by the concept 
of molecular traffic control (MTC) \cite{Dero80,Dero94} which has remained controversial
for a long time. This notion rests on the assumption that reactants and product 
molecules respectively may prefer spatially separated diffusion pathways and thus 
avoid mutual suppression of self-diffusion inside the grain channels.

The necessary (but not sufficient) requirement for the MTC effect, a channel
selectivity of two different species of molecules, has been verified by
means of molecular dynamic (MD) simulations of two-component mixtures
in the zeolite ZSM-5 \cite{Snur97} and relaxation simulations of a
mixture of differently sized molecules (Xe and SF$_6$) in a bimodal
structure possessing dual-sized pores (Boggsite with 10-ring and 12-ring pores)
\cite{Clar00}. Also equilibrium Monte-Carlo simulations demonstrate that the
residence probability in different areas of the intracrystalline pore space
may be notably different for the two components of a binary mixture
\cite{Clar99} and thus provide further support for the notion
of channel selectivity in suitable bimodal channel topologies.

\begin{figure}
\centerline{
\includegraphics[width=4cm]{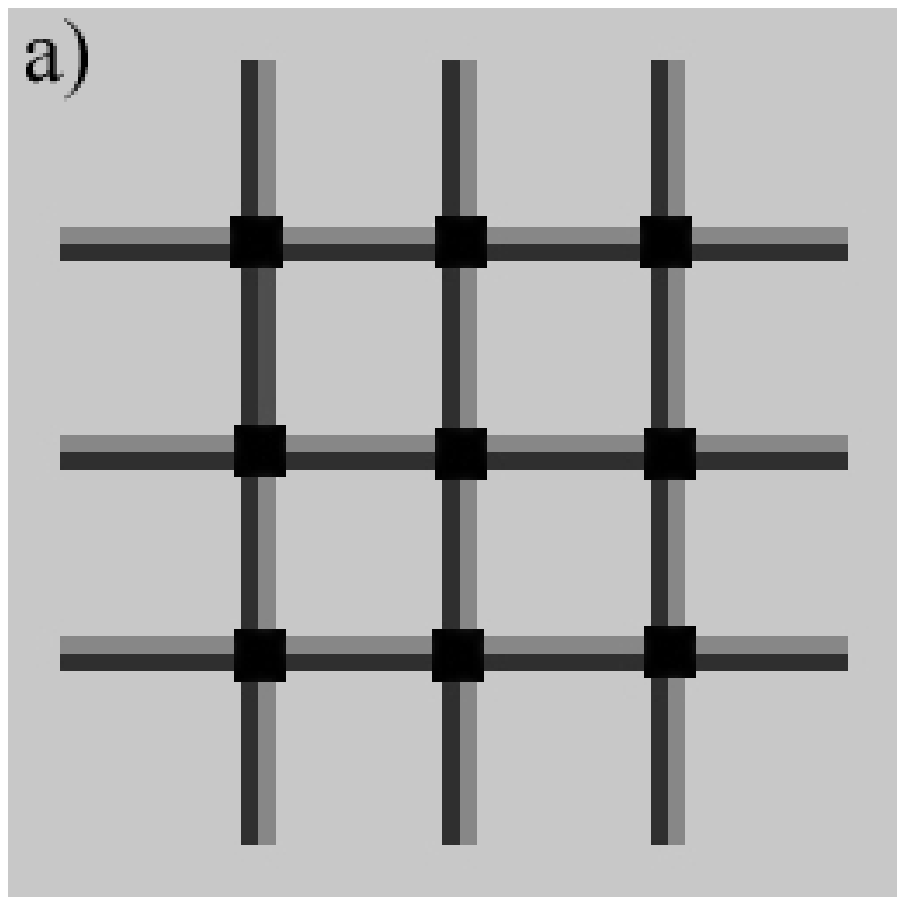}
\includegraphics[width=4cm]{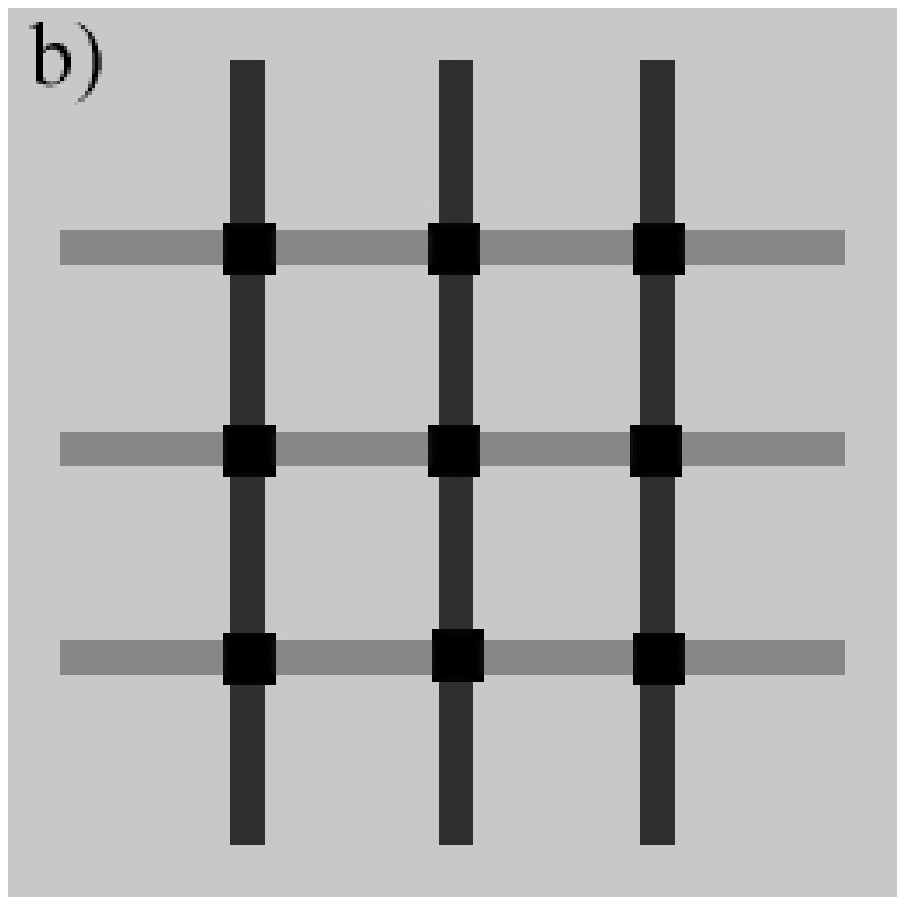}
}
\caption{REF system a), NBK model consisting of perpendicular $\alpha$/$\beta$-channels b)
 - both with channel number of $N=3$. In contrast to the REF case, where we allow both types of particles 
($A$ and $B$ particles) to enter any channel, in the MTC system $A$ particles are carried through the $\alpha$-
channels whereas the $B$ particles diffuse along the $\beta$-
channels. Black squares indicate catalytic sites where a catalytic
transition $A\to B$ takes place.}
\label{systemPics}
\end{figure}

Whether an MTC effect leading to reactivity enhancement actually takes place was
addressed in a series of dynamic Monte Carlo simulations (DMCS) of a stochastic
model system with a network of perpendicular sets of bimodal intersecting channels
and with catalytic sites located at the intersecting pores (NBK model Fig. \ref{systemPics} b)) \cite{Neug00,Karg00,Karg01,Brza05,Brza05-2}. The authors
of these studies found numerically the occurrence of the MTC effect by comparing
the outflow of reaction products in the MTC system with the outflow from
a reference system (REF model Fig. \ref{systemPics} a))  with equal internal and external system parameters, as well as equal internal and optimal external system parameters. The dependency of the 
MTC effect as a function of the system size and catalytic reactivity has been investigated in \cite{Brau03,Brza05}.
The MTC effect is favored by a small number of channels and long channels between intersections, which 
by themselves lead to a very low absolute outflow compared to a similar system 
with shorter channels. For reasonable reactivities and channel lengths the 
MTC effect vanishes inversely proportional to the grain diameter.
An analytical treatment of the master equation for this stochastic many-particle
model revealed the origin of this effect at high reactivities \cite{Brza04}.
It results from an interplay of the long residence time of guest molecules
under single-file conditions with a saturation effect that leads to a depletion
of the bulk of the crystallite. The extension of the NBK model to three dimensions \cite{Brza05-2} leads to 
similar results with the conclusion that MTC may enhance significantly the 
effective reactivity in zeolitic nanoparticles with suitable binary channel 
systems and thus may be of practical relevance in applications. However, no MTC effect
with sufficient absolute effective reactivity can be expected with NBK channel
topology for large grains of at present commercially used size.

\begin{figure}
\centerline{
\includegraphics[width=8cm]{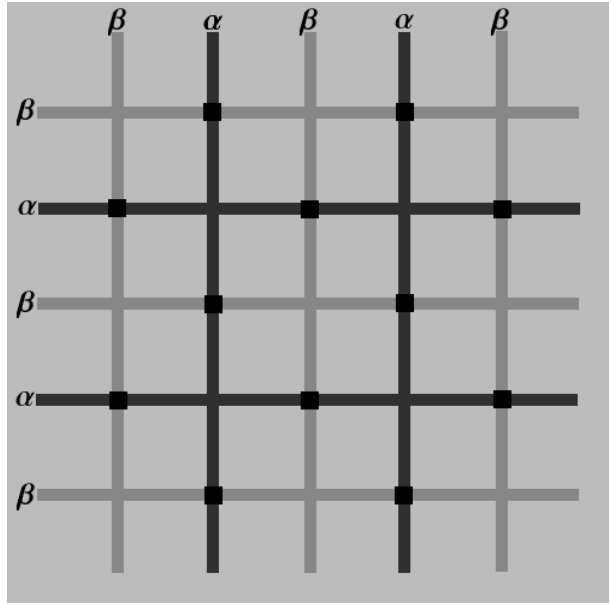}}
\caption{
  MTC system with alternating $\alpha$/$\beta$-channels ($N=5$).}
\label{systemPic2}
\end{figure}

Here we address this question by investigating the MTC effect for a
novel channel-topology (Fig. \ref{systemPic2}) consisting of alternating
$\alpha$-and $\beta$-channels. With a view on applications 
we focus on small channel length. Moreover, for the same reason
we determine the MTC effect by making a comparison with the reference system
using the same set of fixed internal (material-dependent) parameters, but (unlike
in previous studies \cite{Karg01,Brau03,Brza04}) for each
case (MTC and REF resp.) different optimal external (operation-dependent)
parameters which one would try to implement in an industrially relevant
process. This may be of interest as since the first successful
synthesis of mesoporous MCM-41 nanoparticles \cite{Beck92} there has been
intense research activity in the design and synthesis of structured mesoporous
solids with a controlled pore size. In particular, synthesis of bimodal nanostructures
with independently controlled small and large mesopore sizes has become
feasible \cite{Sun03}. It will transpire that a signifcant MTC effect (reactivity
enhancement up to $65\%$) occurs in our model system even for small channel length at
realistic intermediate reaction rates of the catalyst.

\section{The Model}
We consider a lattice model with a quadratic array 
of $N\times N$ channels (Fig. \ref{systemPic2}) which is a measure 
of the grain size of the crystallite. Each channel has $L$ sites
between the intersection points where the irreversible catalytic process 
$A\to B$ takes place. We assume the boundary channels of the grain to be connected 
to the surrounding gas phase, modelled by extra reservoir sites of constant $A$ 
particle density $\rho$. We assume the reaction products $B$ which 
leave the crystallite to be removed immediately from the gas phase such that the 
density of product particles in the reservoir is always 0. Short-range interaction between
particles inside the narrow pores is described by an idealized hard-core
repulsion which forbids double occupancy of lattice sites.

The underlying dynamics are stochastic. We work in a continuous-time description
where the transition probabilities become transition rates and no multiple
transitions occur at the same infinitesimal time unit. Each elementary transition
between microscopic configurations of the system takes place randomly with an
exponential waiting-time distribution. Diffusion is modelled by jump processes
between neighbouring lattice sites. $D$ is the elementary (attempt) rate of
hopping and is assumed to be the same for both species $A,B$ of particles.
In the absence of other particles $D$ is the self-diffusion coefficient for 
the mean-square displacement along a channel. If a neighboring site is occupied by a 
particle then a hopping attempt is rejected (single-file effect). The dynamics inside 
a channel are thus given by the symmetric exclusion process \cite{Spit70,Spoh83,vanB83,Schu94} 
which is well-studied in the probabilistic \cite{Ligg99} and statistical mechanics
literature \cite{Schu01}. The self-diffusion along a channel is anomalous, the
effective diffusion rate between intersection points decays asymptotically
as $1/L$, see \cite{vanB83} and references therein.

At the intersections the reaction $A\to B$ occurs with a reaction rate $c$.
This reaction rate influences, but is distinct from, the effective grain
reactivity which is largely determined by the residence time of guest
molecules inside the grain which under single-file conditions grows in the
reference system with the third power of the channel length $L$
\cite{Rode99}. At the boundary sites, particles jump into the reservoir with
a rate $D(1-\rho_A-\rho_B)$ in the general case. Correspondingly,
particles are injected into the grain with rates $D\rho_{A,B}$ respectively.
As discussed above here we consider only $\rho_A=\rho$, $\rho_B=0$.

For the REF system reactants as well as product particles are allowed to enter and leave both types 
of channels. In case of MTC reactant(product) particles will enter $\alpha$($\beta$)-channels only,
mimicking complete channel selectivity. Therefore all channel segments carry
only one type of particles in the MTC case. For the boundary channels
complete selectivity implies that $\alpha$-channels are effectively
described by connection with an $A$-reservoir of density $\rho_A=\rho$
(particles of type $B$ do not block the boundary sites of $\alpha$-channels)
and $\beta$-channels  are effectively described by connection with a
$B$-reservoir of density $\rho_B=0$, respectively.
(particles of type $A$ do not block the boundary sites of $\beta$-channels.)
This stochastic dynamics, which is a Markov process, fully defines our MTC 
model.

In both cases, MTC and REF, the external concentration gradient 
between reactant and product reservoir densities induces a particle current inside the
grain which drives the system in a stationary nonequilibrium state.
For this reason there is no Gibbs measure and equilibrium
Monte-Carlo algorithms cannot be applied for determining steady-state
properties. Instead we use dynamic Monte-Carlo simulation (DMCS)
with random sequential update. This ensures that the simulation algorithm yields 
the correct stationary distribution of the model.

\section{Monte Carlo results}

Anticipating concentration gradients between intersection points we expect due 
to the exclusion dynamics linear density profiles within the channel
segments \cite{Spoh83,Schu01,Brza04}, the slope and hence the current
being inversely proportional to the number of lattice sites $L$.
The total output current $j$ of the product particles, defined as the number of
$B$ particles leaving the grain per time unit in the stationary state, is the main
quantity of interest. It determines the effective reactivity of the grain.

\begin{figure}
\centerline{
  \includegraphics[width=5cm,angle=270]{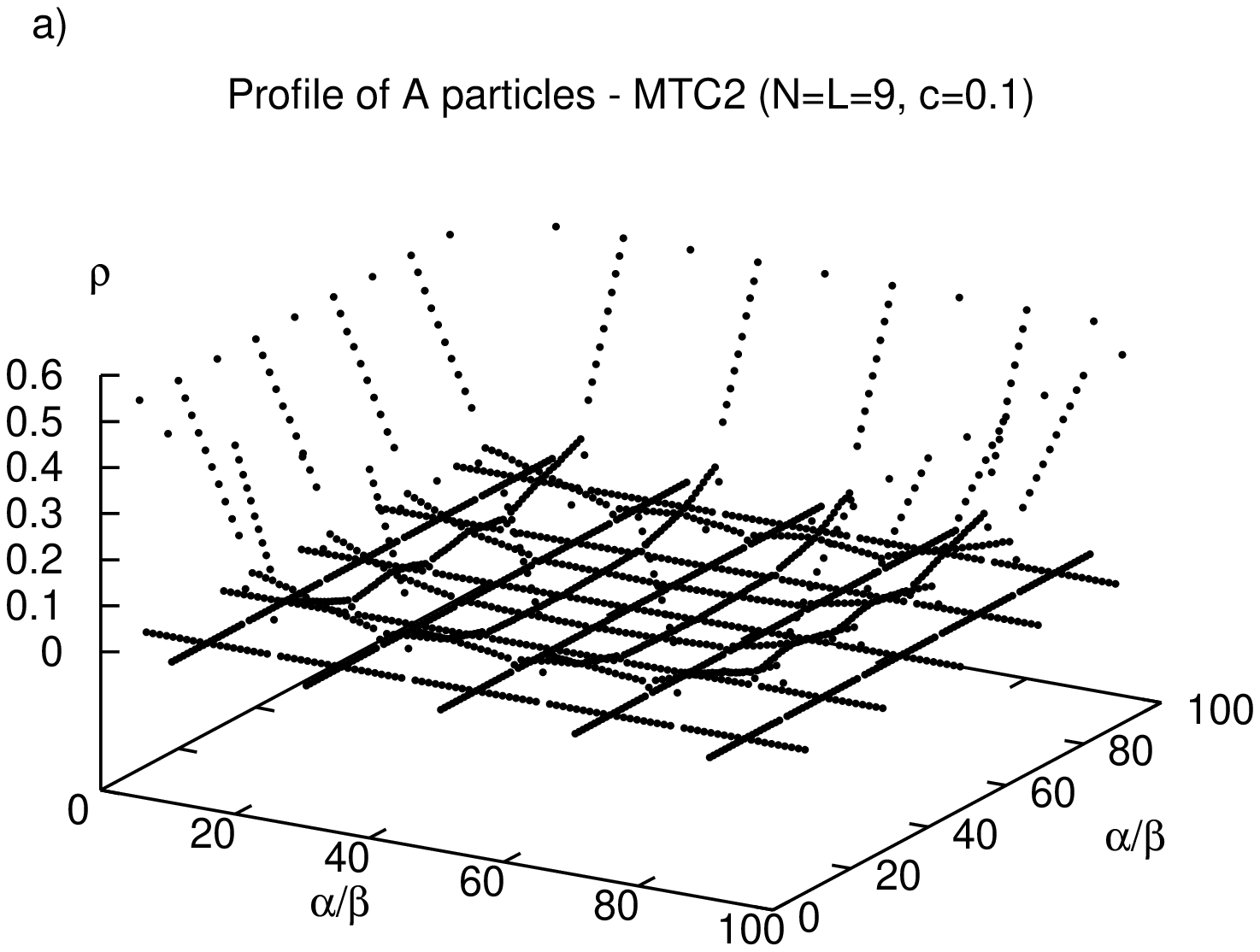}
  \includegraphics[width=5cm,angle=270]{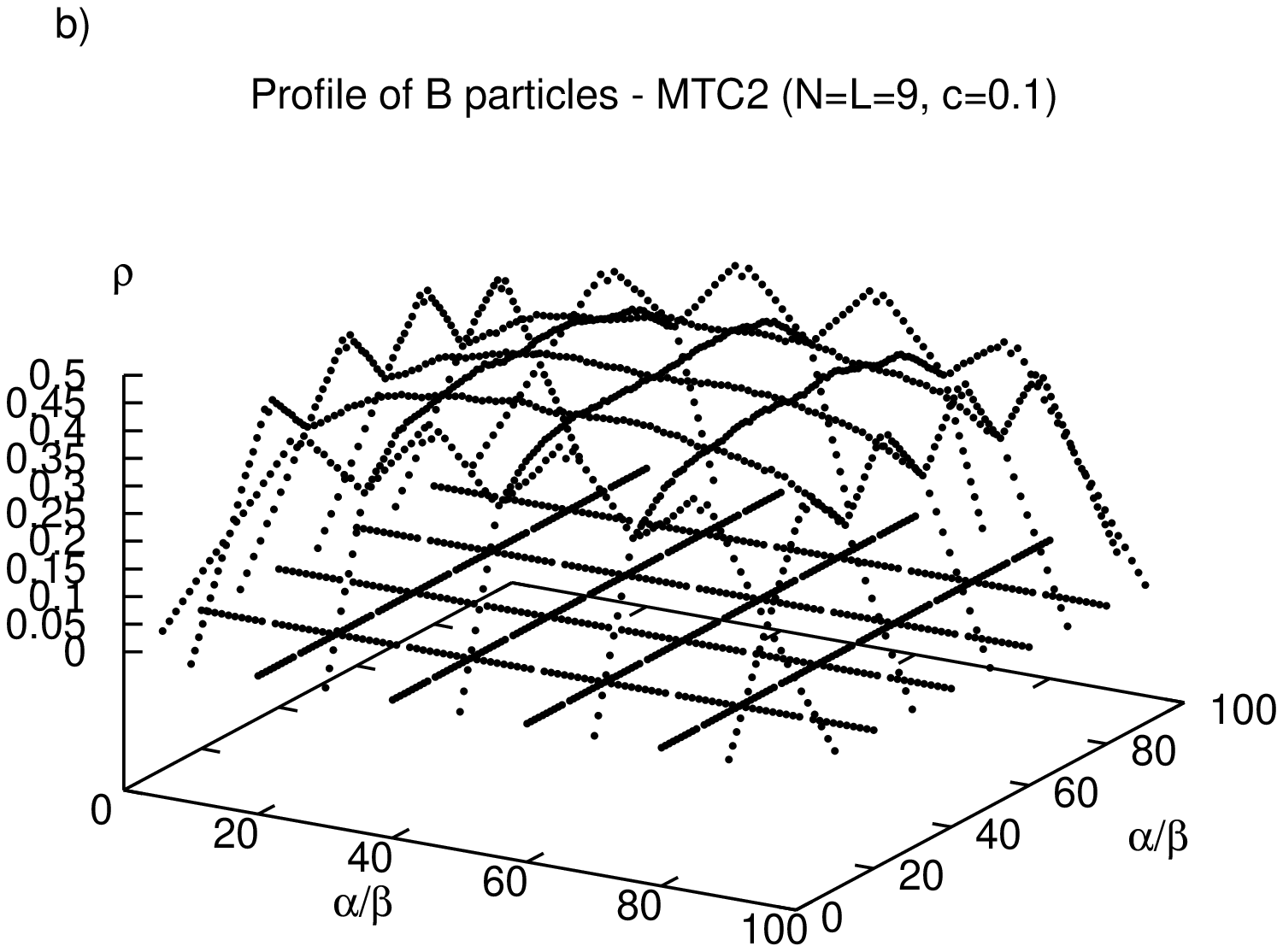}}
\caption{A/B particle profile (a)/b)) for the MTC system of $N=L=9$ and reactivity $c=0.1$.}
\label{profiles}
\end{figure}

\begin{figure}
\centerline{
  \includegraphics[height=7cm,angle=270]{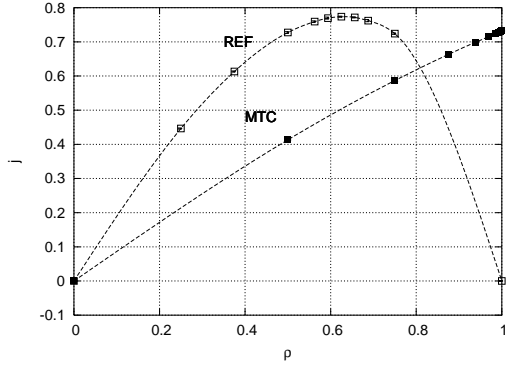}}
\caption{$j_{REF}$ (solid symbols) and $j_{MTC}$ (open symbols) as a function
of the reservoir density for a system with  $N=5$, $L=1$, $c=0.1$.}
\label{rhostar}
\end{figure}

\begin{figure}
\centerline{
  \includegraphics[width=5cm,angle=270]{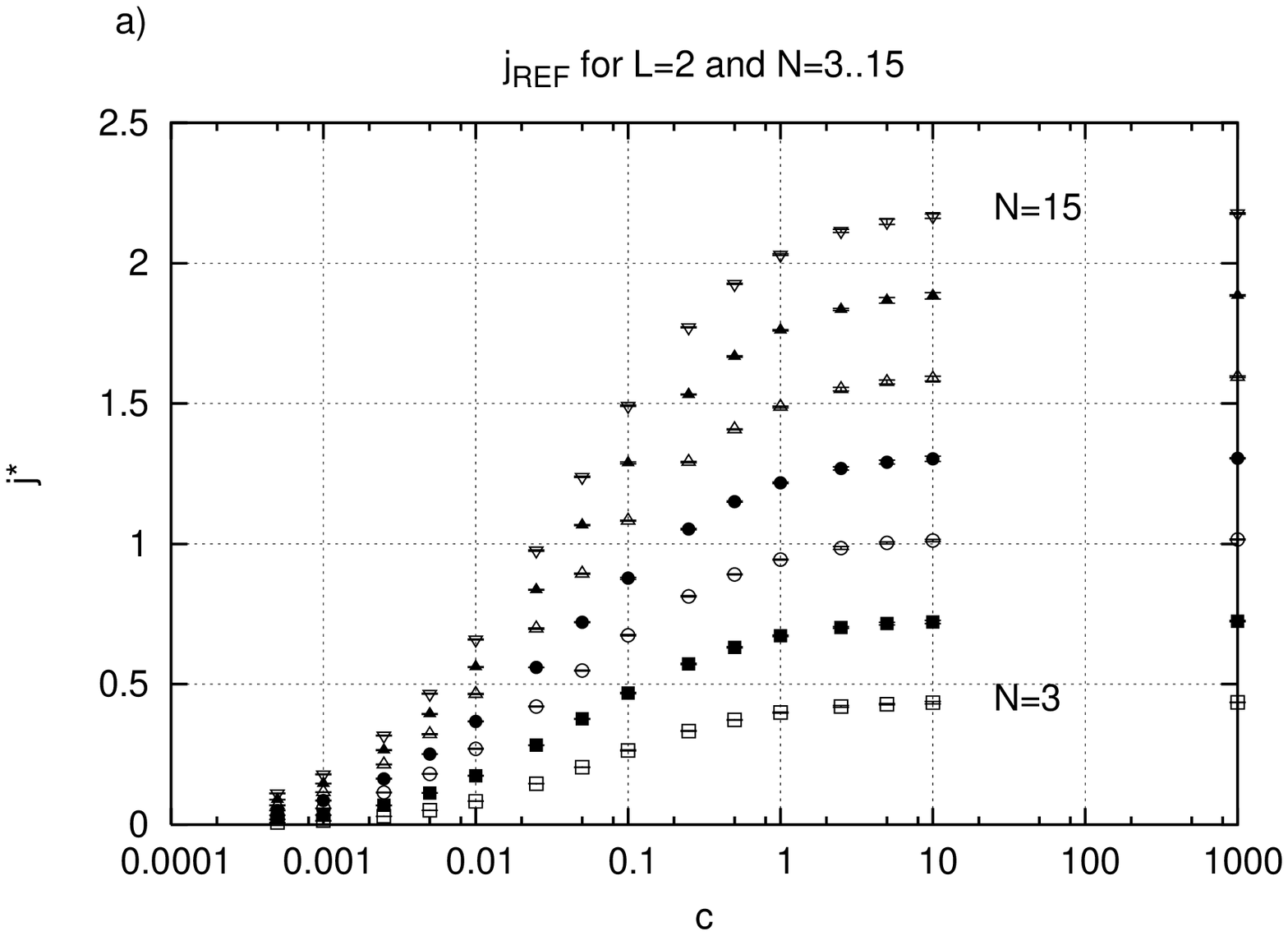}
  \includegraphics[width=5cm,angle=270]{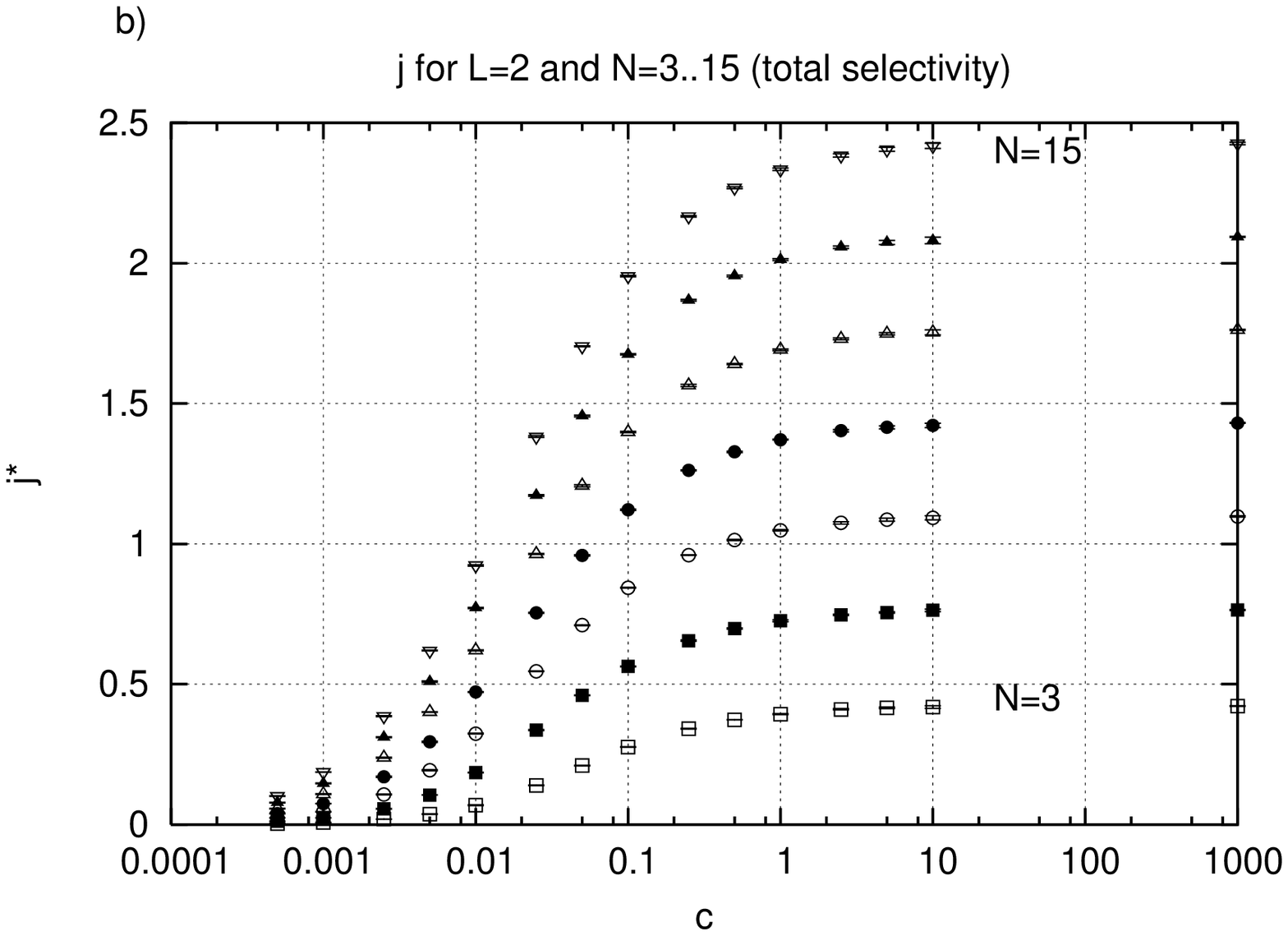}
}
\caption{Maximal currents $j^*$ over reactivity $c$ for the REF system a) and
MTC system b) for different $N$.}
\label{currents}
\end{figure}

Fig. \ref{profiles} shows the particle densities for the reactants a) and product particles 
b) in the stationary regime for
the MTC system $N=L=9$ and reactivity $c=0.1$. Every other boundary-channel segment 
contributes to the output, therefore we expect no efficiency loss with increasing $N$.
We are particularly interested in studying the system in its maximal current state 
for given reactivity $c$ and size constants $N$, $L$, which are
intrinsic material properties of the grain. The reactant particle reservoir
density $\rho$, determined by the density in the gas phase, can be tuned in
a possible experimental setting. Let us therefore denote the reservoir density 
which maximizes the output current with $\rho^*$ and the maximal current with $j^*$.
We iteratively approach the maximal current by a set of 9 datapoints. 
Fig. \ref{rhostar} shows $j$ as a function of $\rho$ for both a MTC and REF system 
of $N=5$, $L=1$ and reactivity $c=0.01$. The "best" datapoint has been chosen in order 
to approximate the maximum. Statistical errors are displayed, they are, however, mostly 
within symbol size. We note that for the MTC system the maximal current is attuned for
$\rho^*=1$. REF and MTC system show a similar dependence of the maximal current as
a function of the reactivity (Fig. \ref{currents}). In contrast to the NBK topology 
\cite{Brza05}, $j^*(c)$ is a monotonically increasing function and the highest product particle
output is achieved for infinite reactivity. In this case any transition A$\to$B transpires
close to the surface and hence, from an applied point of view one would be interested in
various small grains rather than view big ones.

In order to measure the efficiency of an MTC system over the associated REF system
we define the efficiency ratio
\begin{align}
R(c,N,L)=\frac{j_{MTC}^*}{j_{REF}^*}
\end{align}
which is a function of the system size $N$, $L$ and reactivity $c$.

\begin{figure}
\centerline{
  \includegraphics[width=5cm,angle=270]{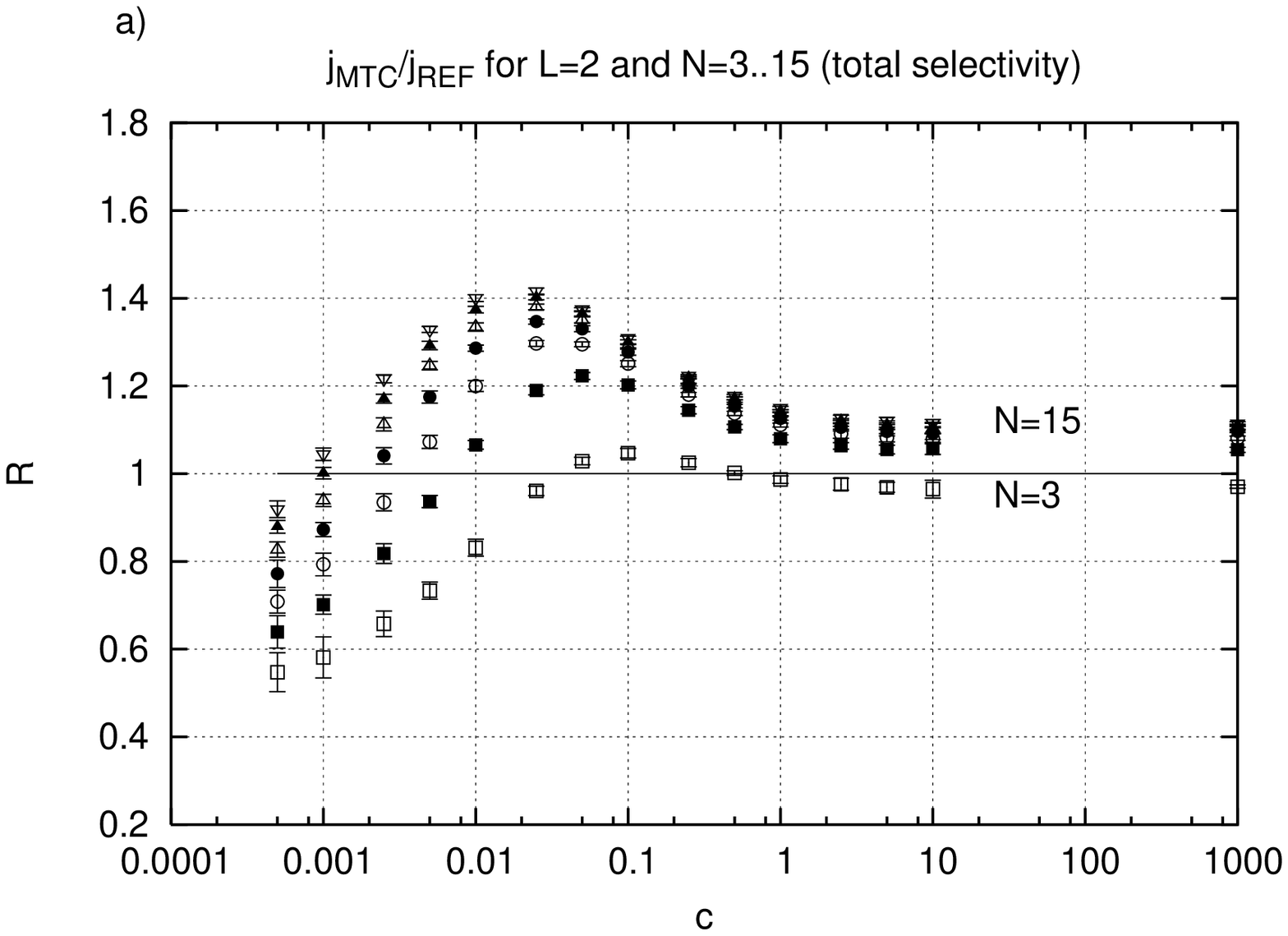}
  \includegraphics[width=5cm,angle=270]{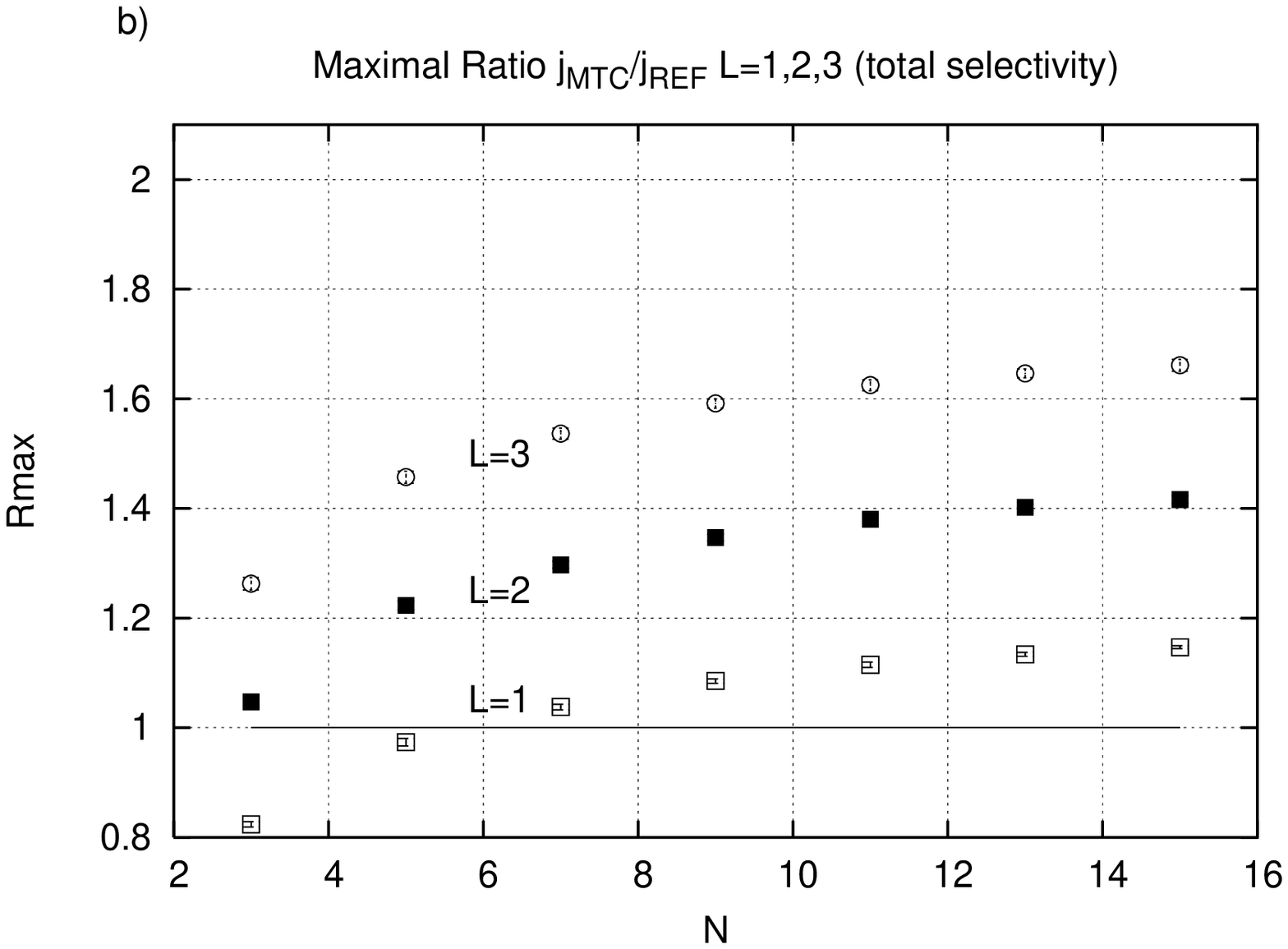}
}
\caption{a) Ratio $R(c)$ for different number of channels $N$ and $L=2$. 
b) Maximal ratio $R^*$ for different $L$}
\label{RMTC}
\end{figure}

Fig. \ref{RMTC} a) shows the measured ratio $R$ for a large range of 
reactivities. We plotted systems with $L=2$ and different $N$. We find 
an MTC effect over a large range of reactivities. There is a small tendency
of growing efficiency with increasing $N$. This is highly desirable
since an enhancement of efficiency due to MTC is thus not restricted to nano-scale 
crystallites. Compared to former studies of MTC systems, this is a qualitatively
new property and results from the fact that the number of output ($\beta$-channels) 
scales linearly with growing system size $N$. In general, compared to the NBK model,
the efficiency ratio keeps quite unchanged in the regime of simulated reactivities. 
There is no sharp decrease for large reactivities. Nevertheless, we notice an optimal 
reactivity $c^*$ for which $R$ becomes maximal. This value is denoted 
by $R^*$ and plotted in Fig. \ref{RMTC} b) for increasing $N$ and different $L$. 
The MTC effect is present for any $L$ and our model exhibits an enhancement of the 
effective reactivity of up to $20\%$ for (extremely) short channel length ($L=1$) and
up to $65\%$ for $L=3$. 

\begin{figure}
\centerline{
  \includegraphics[width=7cm,angle=270]{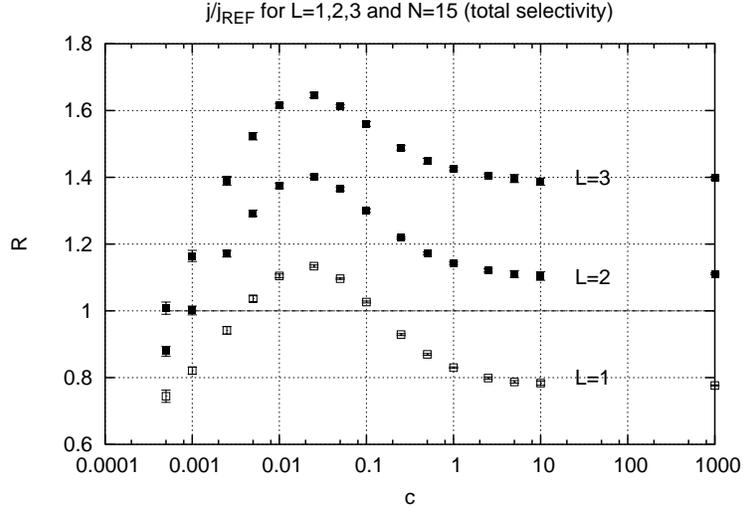}
}
\caption{Dependence of the output ratio on the channel length $L$.}
\label{increasingL}
\end{figure}

As expected from previous results \cite{Brau03,Brza04}  the MTC effect
is seen to increase with increasing $L$ (Fig. \ref{increasingL}). This follows 
from theoretical studies of single-file systems which predict that the mean travelling time of a 
product molecule through a channel of length $L$ is proportional
to $L^2$ in the MTC case as in ordinary diffusion, but proportional to $L^3$
in the REF case due to mutual blockage. Hence the current is proportional to
$1/L$ in an MTC system, but proportional to $1/L^2$ in a REF system. This
holds for all values of the parameters and hence, for sufficiently large $L$,
the MTC system becomes more efficient.

\section{The high-reactivity case}
\begin{figure}
\centerline{
  \includegraphics[width=5cm,angle=270]{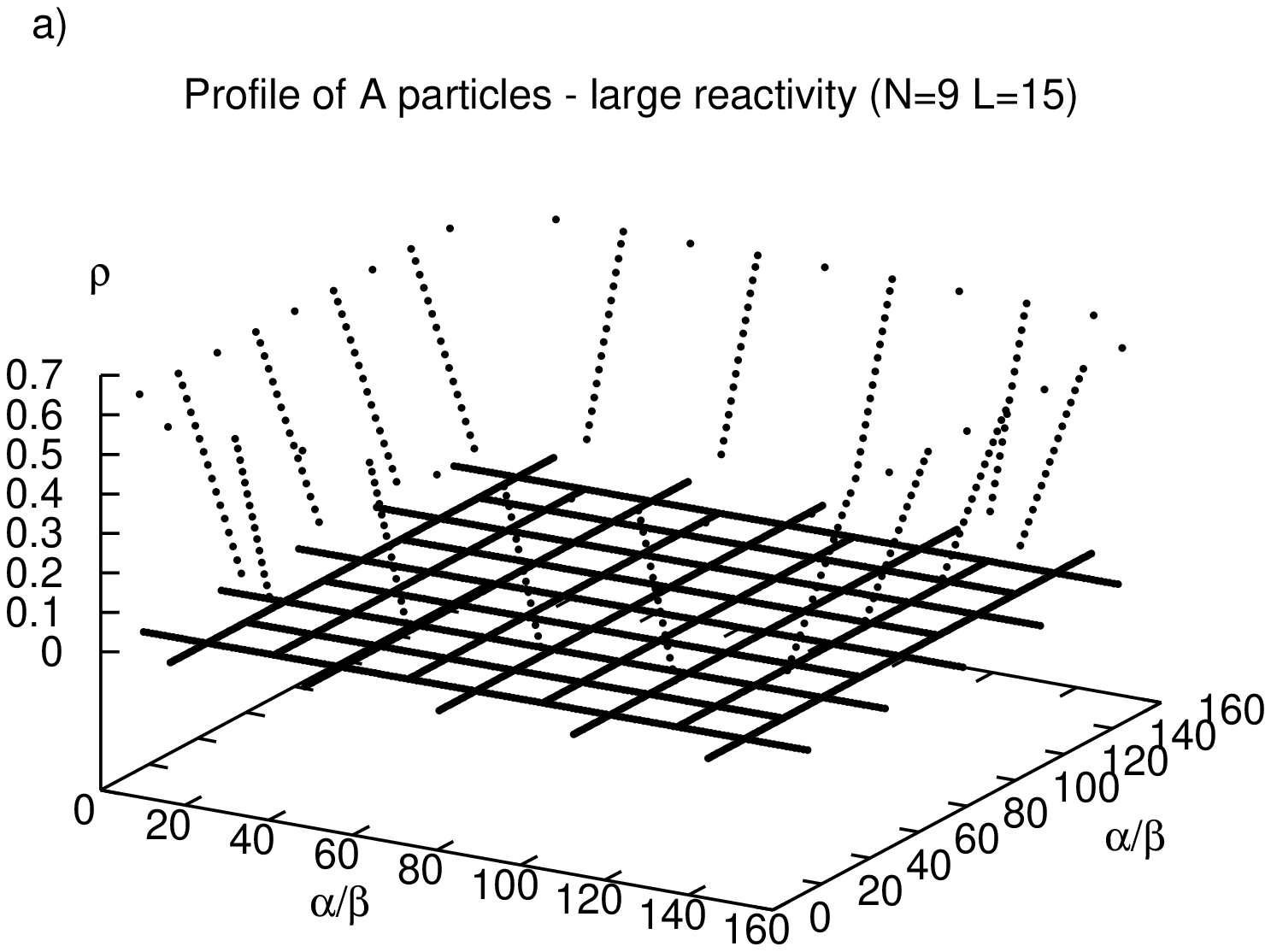}
  \includegraphics[width=5cm,angle=270]{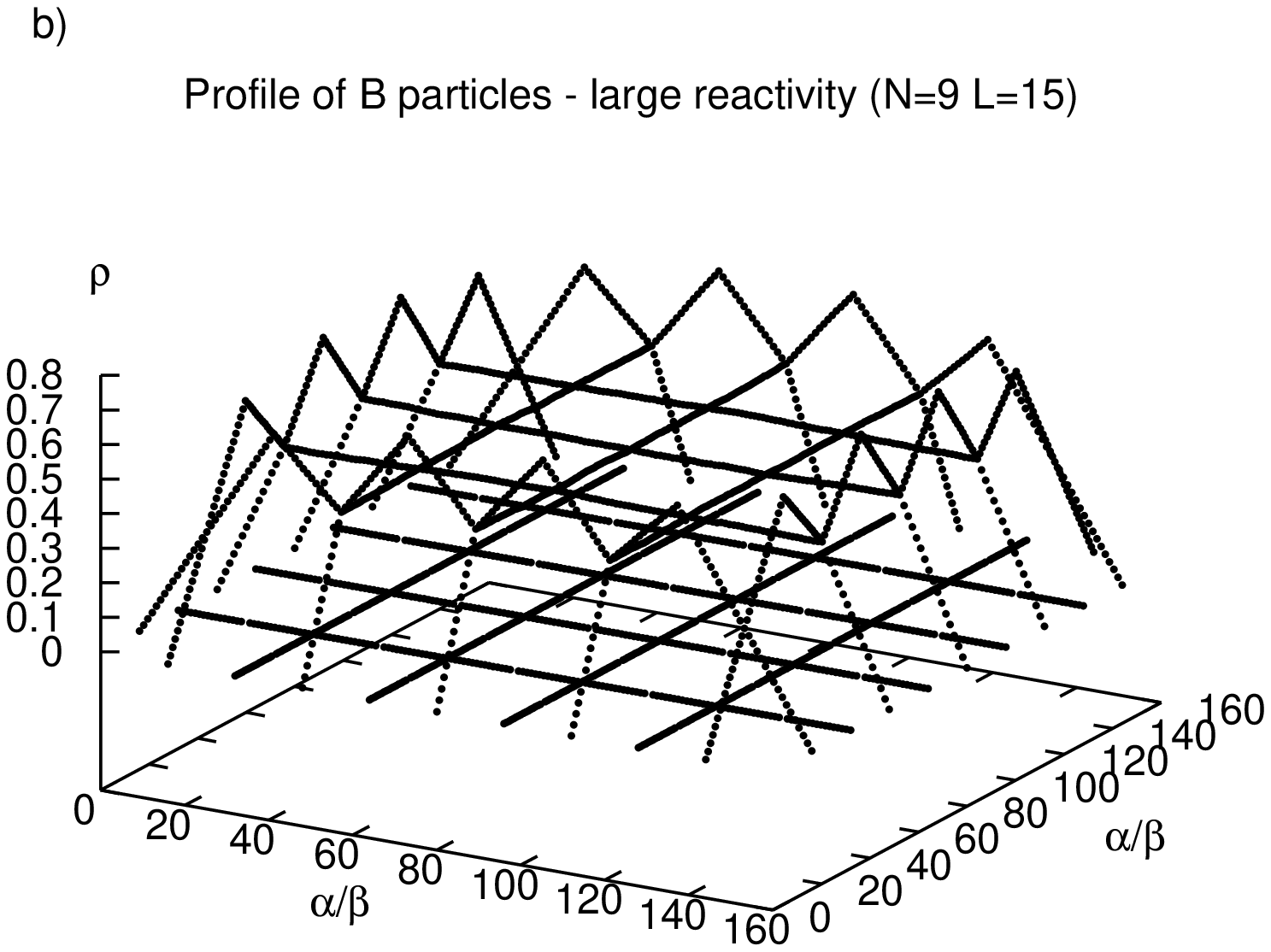}}
\caption{A/B particle profile (a)/b)) for a system of $N=9$, $L=15$ and
reactivity $c=1000$.}
\label{profilesFast}
\end{figure}

\begin{figure}
\centerline{\includegraphics[width=11cm]{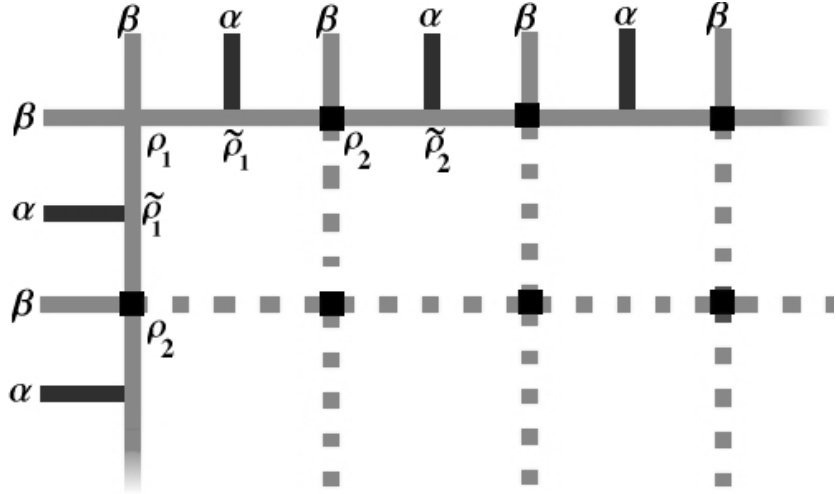}}
\caption{Corner of the reduced system for the case of fast reactivity.}
\label{fastReactivity}
\end{figure}

The penetration depth of reactant particles into the system is related to 
the reactivity $c$. In the fast-reactivity case transitions and 
particle transport take place close to the grain-surface. Fig. \ref{profilesFast}
shows the profile of the reactant particles a) and product particles b) for a system of $N=9$ channels, 
$L=15$ and fast reactivity ($c=1000$). Notice that almost no reactants reach
the bulk of the system since reactants entering the system are converted 
very likely into product particles at the first intersection. Hence, a 
creation of product particles occurs only on the outer $\beta$-channels, 
which leads to equilibration of the product particles in the bulk of the system. This 
picture is supported by simulation (Fig. \ref{profilesFast}) and suggests to define 
a reduced system, illustrated in Fig. \ref{fastReactivity}, for the following theoretical 
analysis. 

Here, $\alpha$-segments carrying no particles are suppressed, $\beta$-segments 
in equilibrium are represented by dotted lines. The density profile in a channel 
segment enclosed by two intersections or an intersection and boundary site is linear 
due to pure exclusion dynamics \cite{Schu94}. This allows us to reduce our effort to the 
computation of the intersection densities. The intersection densities are calculate 
from the Master equation. The output of product particles, i.e. the number of $B$'s leaving the 
system, is given by the currents $j_x^B$ of the $\beta$-segments connected to the reservoir. 
The calculation is similar to \cite{Brza04} and we identify 
\begin{align}
\label{currentB}
j_x^B&=D\frac{\rho_x}{L+1}.
\end{align}
Here, $\rho_x$ is the product particle density at intersection $x$ connected to 
the $\beta$-Segment. The derivation of the currents involves a simplification. Within a mean field 
approximation, joint probabilities $\lav x y \rav$ of particle 
numbers between an intersection and its adjacent sites can be replaced by the 
product of the individual averages $\lav x \rav\lav y \rav$. This well controlled approximation 
is, in fact, necessary only in the vicinity of intersections, since any joint probability vanishes
in the bulk of the channels \cite{Schu94}. The expected relative error is of the order of $1/L$
\cite{Spoh83,Schu01}. Reactant particles diffuse into the system through $\alpha$-segments 
which are bounded at one side by the reservoir (density $\rho_A$) and at the other side 
by intersection $\tilde{\rho}_x$ with current
\begin{align}
j_x^A&=D\frac{\rho_A\left(\tilde{\rho}_x-1 \right)}{L\left( \tilde{\rho}_x-1 \right)-1}.
\end{align}
Finally, the intrinsic currents, determined by two adjacent intersections follow
to
\begin{align}
\overleftarrow{j}_x^B&=D\frac{\tilde{\rho}_x-\rho_x}{L+1}\\
\overrightarrow{j}_x^B&=D\frac{\tilde{\rho}_x-\rho_{x+1}}{L+1}.
\end{align}
The arrows are not supposed to indicate vectors but to distinguishe the direction of 
currents in order to comply with the zigzag profile shown in Fig.\ref{profilesFast}.

\begin{figure}
\centerline{
\includegraphics[width=7cm,angle=270]{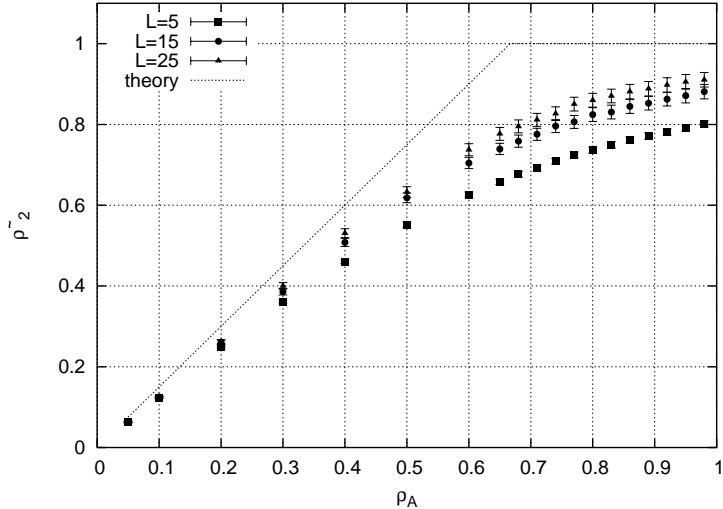}
}
\caption{Density $\tilde{\rho}_2$ for different $L$ as a function of the reservoir density.
The theoretical density is indicated by the dotted line. $N=9$}
\label{profQual}
\end{figure}

Making use of conservation of currents
\begin{align}
\label{conservation}
j_x^A&=\overleftarrow{j}_x^B+\overrightarrow{j}_x^B\\
j_x^B&=\overleftarrow{j}_x^B+\overrightarrow{j}_{x-1}^B.
\end{align}
we find (for large but finite $L$) two coupled recursion relations for the intersection densities which
determines the density profile. 
\begin{align}
\label{recursions}
\tilde{\rho}_x&=
\begin{cases}
1\\
\frac{1}{2}(\rho_x+\rho_{x+1}+\rho_A)
\end{cases}\\
3\rho_x&=\tilde{\rho}_x+\tilde{\rho}_{x-1}
\end{align}
Let $N_{\alpha}=\lfloor N/2 \rfloor$ be the number of $\alpha$-channels. In order to ensure zero current 
in the equilibrium parts of the $\beta$-channels (Fig. \ref{fastReactivity}), intersections marked 
with a black square need to attain equal densitiy. Hence, in the region $2\le x\le N_{\alpha}$ we assume a constant solution for the densities $\rho_x$ and find $\rho_x \equiv \rho=\rho_A$.
Intersections $\tilde{\rho}_x$ connected to $\alpha$-channels carry a product particle density
being $3/2$ times higher (Fig. \ref{profQual}) and therefore $\tilde{\rho}_x \equiv \tilde{\rho}=3/2\rho_A$. 
Since the particle density can not exceed 1, $\rho$ saturates for a boundary density of $\rho_A=2/3$.
This has an interesting consequence. Changing the boundary density from $\rho_A=2/3$ 
to $\rho_A=1$ merely shifts up the profile of the $\alpha$-channels,
the output, however, keeps unchanged. For the sake of completeness we remark
that for the four corners $2\rho_1=\tilde{\rho}_1$, hence, $3\rho_1=2\rho_A$. \\

We have now determined the profile. Because of \eqref{currentB}
the output of product particles for each $\beta$-channel is independent of the number of channels, hence, the 
total output current scales linearly with $N$.
Figure \ref{profQual} a) shows the profile of intersection $\tilde{\rho}_2$ as a function
of the reservoir density for a system of $N=9$ channels and different $L$. The saturation 
at $\rho_A=2/3$ predicted for $L \to \infty$ (vanishing corrections to mean field theory) becomes more pronounced
for increasing $L$. The theoretical description
is good for small reservoir densities. The largest deviation (about $20\%$ at most) occurs for the predicted 
saturation point, so for practical purposes the theoretical analysis agrees well with 
the simulated data.

\section{MTC Model with size selectivity}
For situations where channel selectivity merely originates from different pore 
diameters we have to modify our model. 
Consider the model shown in Fig. \ref{systemPics} but now, allowing reactant as well as product particles
to diffuse along the $\alpha$-channels (size selectivity). We call them
$\gamma$-channels. Assuming $\beta$-channels to be smaller in diameter than $\alpha$-channels there 
is no reason why particles of type $B$ would avoid to occupy $\alpha$-channels. Of course we
could also consider the situation vice versa, with $\beta$ being larger than $\alpha$.
However, this leads to a reduction of output channels, hence, this situation would be
counterproductive and is not studied here. Overtaking is still prohibited for any type of particles. 
The dynamics explained in the model-section keeps unchanged and similar to arguments
given before we would like to compare with the REF system for 
equal internal parameters ($N, L, c$) but in their maximal current state with respect to
the reservoir density $\rho_A$.\\

\begin{figure}
\centerline{
  \includegraphics[width=5cm,angle=270]{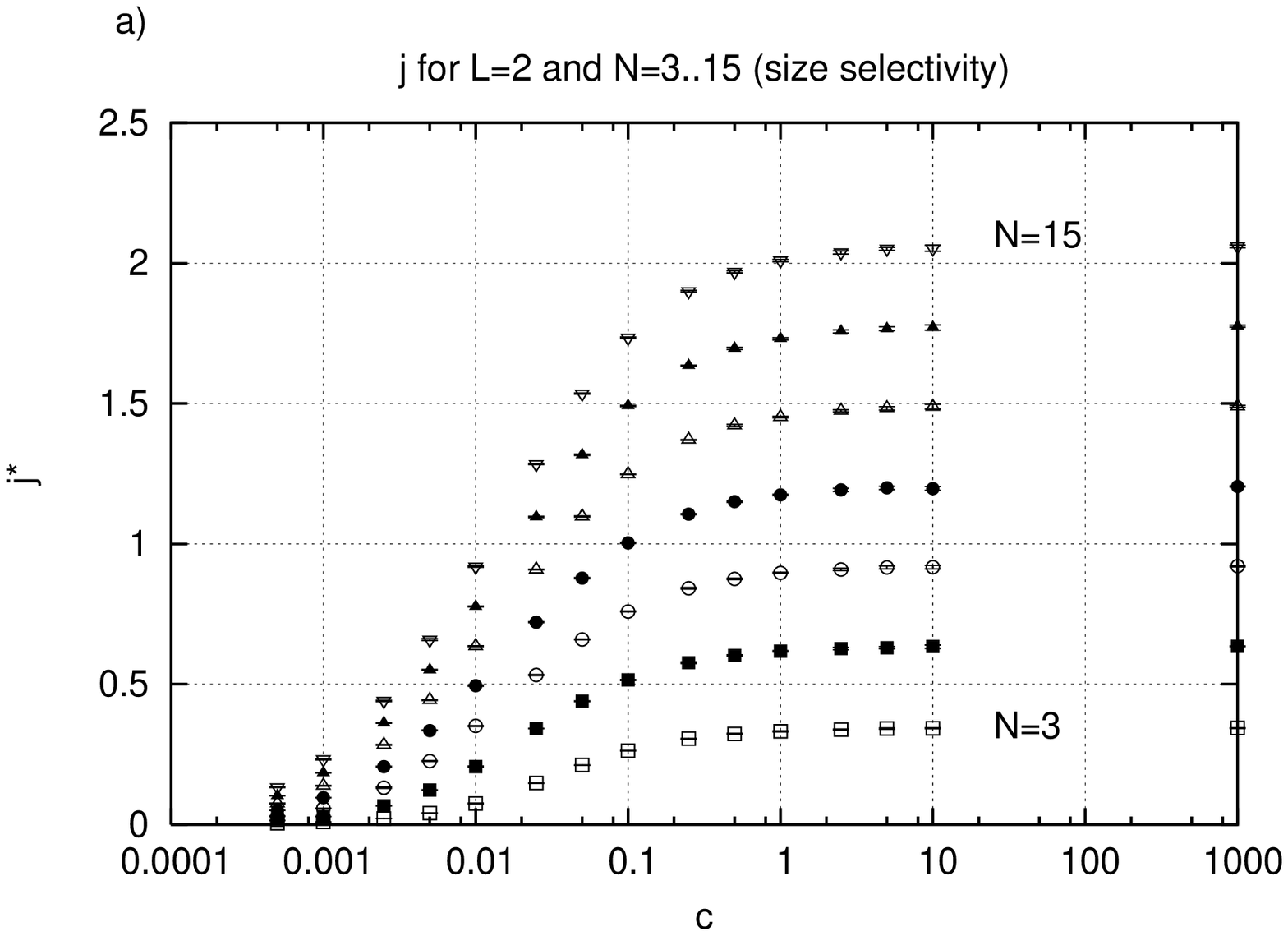}
  \includegraphics[width=5cm,angle=270]{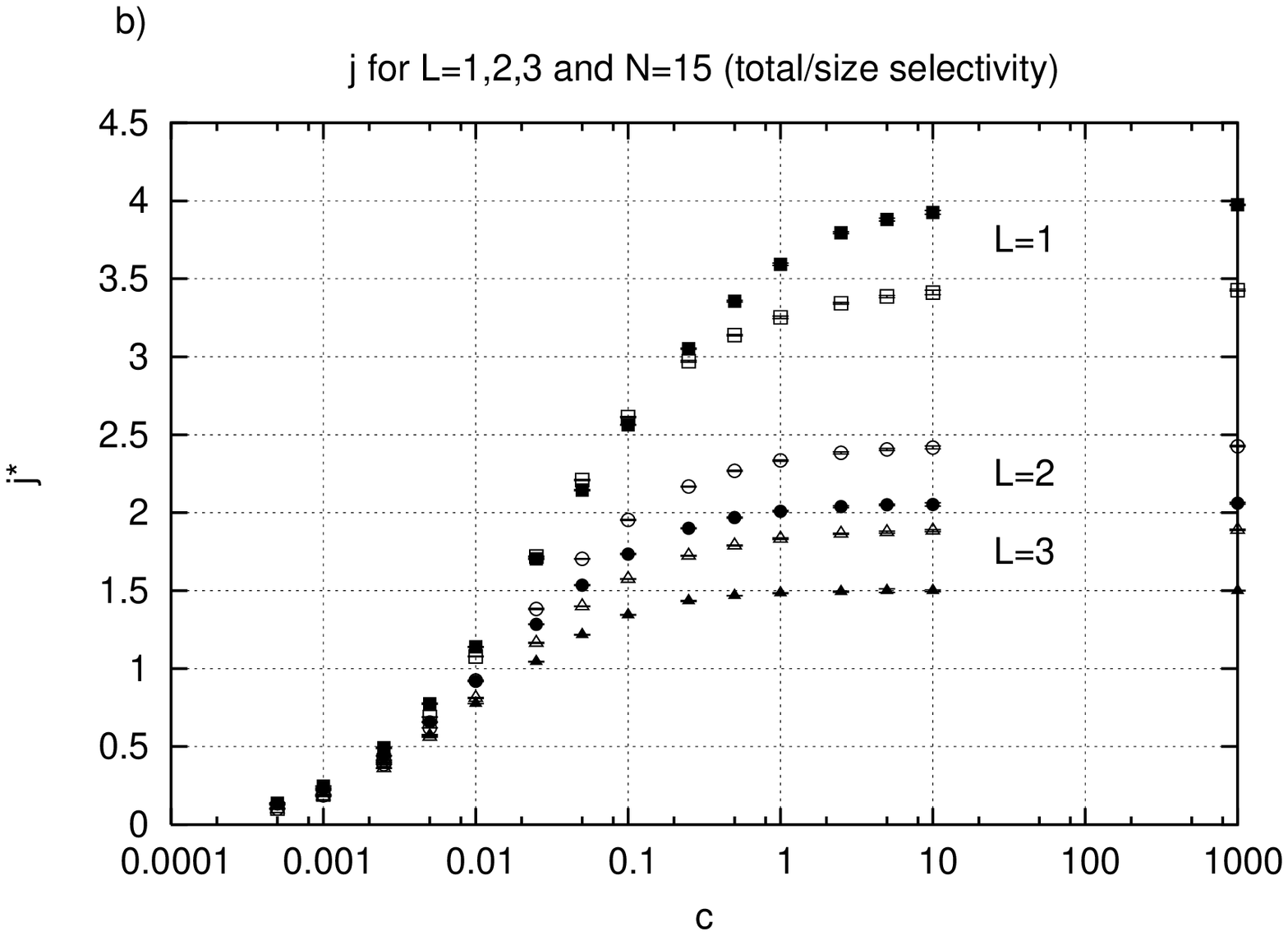}
}
\caption{a) Maximal currents $j^*$ over reactivity $c$ for the size controlled MTC system. b)
Maximal Currents for the size controlled system (solid symbols) and the system of total
selectivity (open symbols).}
\label{gammacurrents}
\end{figure}

\begin{figure}
\centerline{
\includegraphics[width=5cm,angle=270]{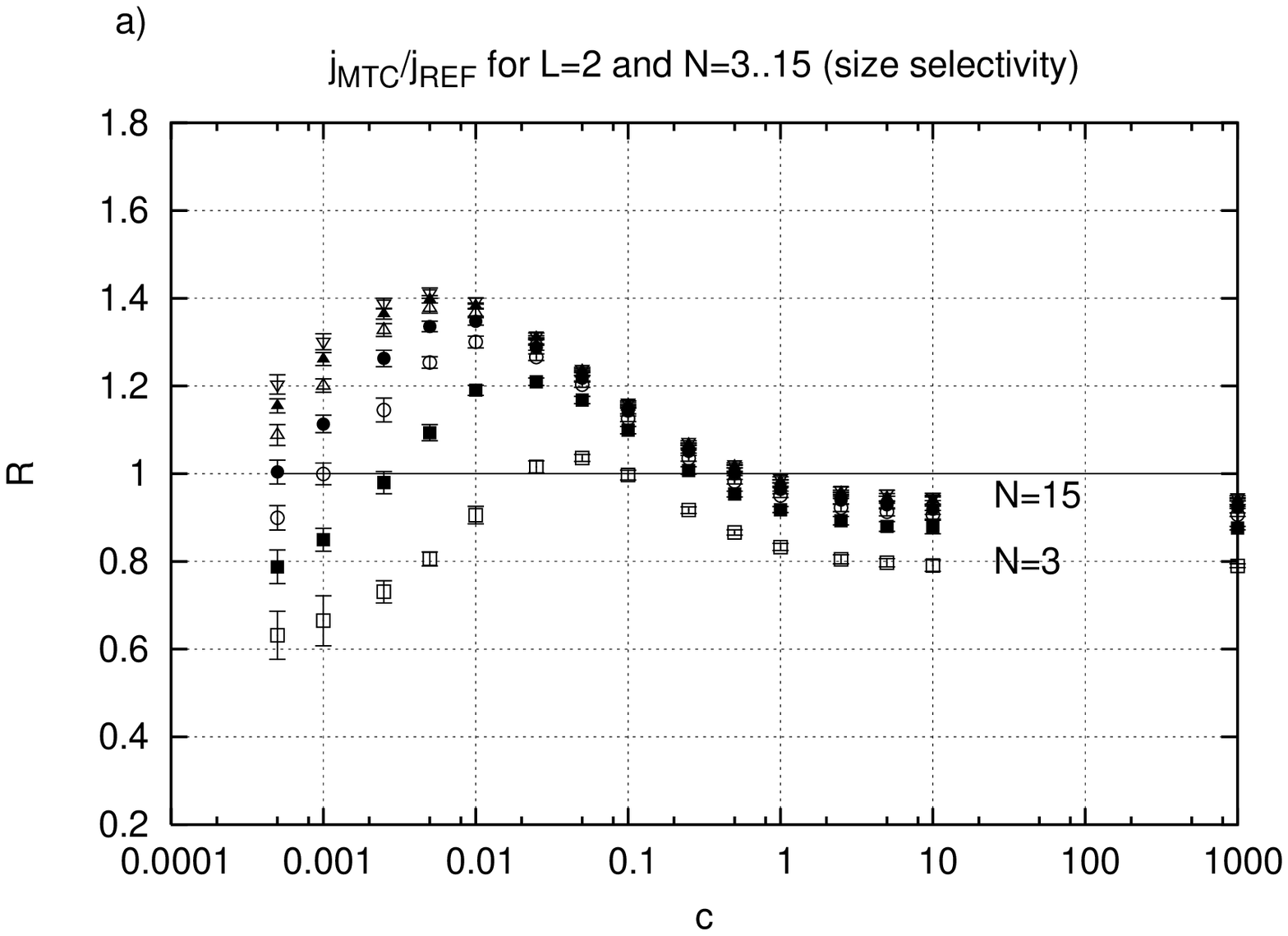}
\includegraphics[width=5cm,angle=270]{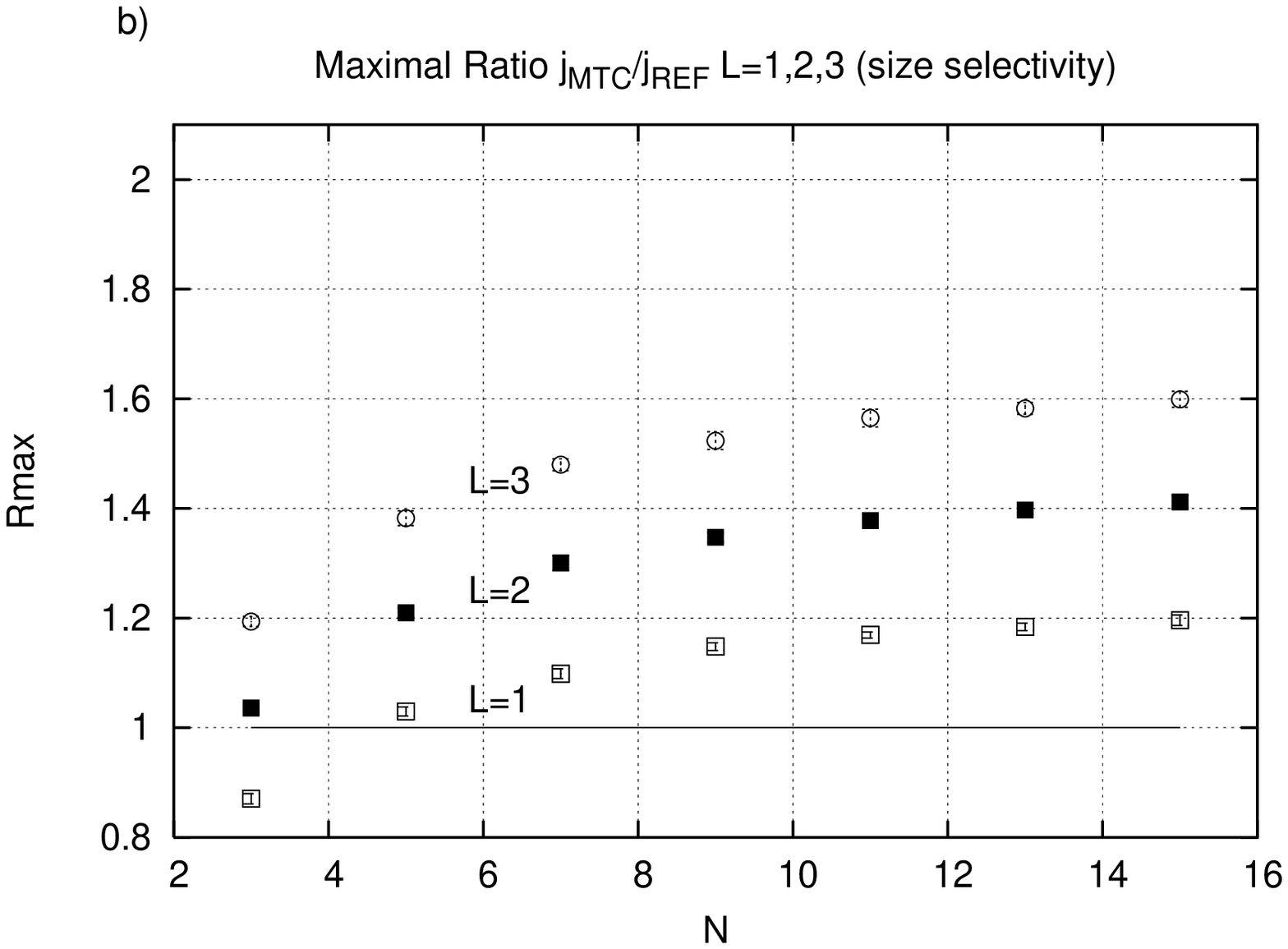}}
\caption{Left: Ratio $R(c)$ for different number of channels $N$ and $L=2$. 
b) Maximal ratio $R^*$ for different $L$. (size selectivity)}
\label{gammaRMTC}
\end{figure}

Comparing the currents of the size-controlled system (Fig. \ref{gammacurrents} a)) with
the model of total selectivity (Fig. \ref{currents} a)) we find a similar 
qualitative behaviour. There is an interesting observation, that the size-controlled model
performes a higher output compared to the model of total selectivity for $L=1$, however, for
larger $L$ total selectivity becomes more and more efficient (Fig. \ref{gammacurrents} b)). This is
true for high enough reactivities. Both models maximize their output for infinite reactivity,
hence, both models are surface active in their state of maximal performance, since
for large reactivities the interior becomes rather inactive due to lacking particles of type $A$. 
The efficiency $R$ is plotted in Fig. \ref{gammaRMTC} a) for $L=2$. Similar to 
the case of total selectivity (Fig. \ref{RMTC} a)) $R$ undergoes a maximum for some
intermediate $c$, however, we observe a shift to smaller reactivities. Plotting
the maximal efficiency for increasing $N$ (Fig. \ref{RMTC} b)) proves the fact that
the MTC effect is present for any $L$.

Ack

\section{Conclusion}
Our simulations and analytical results describe the MTC effect
quantitatively over a wide range of parameters for a model of
alternating $\alpha$- and $\beta$-channels and novel channel topology. 
The output current scales linearly with the number of channels and, hence, the 
efficiency ratio (compared with a topologically and structurally similar 
reference system without MTC) does not sharply decrease with increasing system size.
This new insight is encouraging as it allows for the 
MTC effect in experimental
settings not only for nanoscale systems. This suggests that MTC may 
enhance significantly the effective reactivity in zeolitic
particles with suitable binary channel systems and thus may
be of practical relevance in applications. Introducing a model
of size selectivity where reactant as well as product particles are allowed
to enter $\alpha$-channels reveals only minor differences in the observed behaviour.
For the important case of small channel segments, size selectivity 
seems to be slightly more efficient compared to total selectivity.

Acknowledgement: Financial support by the Deutsche Forschungsgemeinschaft is 
gratefully acknowledged. We also thank Jörg Kärger and Peter Bräuer for useful 
discussions.

\end{document}